\documentclass[prd,nofootinbib,floats,superscriptaddress,eqsecnum,tightenlines,11pt]{revtex4}

\usepackage{hyperref}
\usepackage{graphicx}
\usepackage{amsmath,amssymb,amsfonts,amsthm,latexsym,stmaryrd}
\usepackage{marginnote}
\usepackage{color}

\def\be{\begin{equation}}
\def\ee{\end{equation}}
\def\ba{\begin{eqnarray}}
\def\ea{\end{eqnarray}}
\def\bas{\begin{subequations}\begin{eqnarray}}
\def\eas{\end{eqnarray}\end{subequations}}

\def\tr{\text{tr}}

\def\SU{\text{SU}}

\def\su{\mathfrak{su}}

\begin{document}

\title{Loop Quantum Cosmology with Complex Ashtekar Variables}

\author{Jibril Ben Achour }
\email{benachou@apc.univ-paris7.fr}
\affiliation{Laboratoire APC -- Astroparticule et Cosmologie, Universit\'e Paris Diderot Paris 7, 75013 Paris, France}

\author{Julien Grain}
\email{julien.grain@ias.u-psud.fr}
\affiliation{%
CNRS, Institut d'Astrophysique Spatiale, UMR8617,Orsay, France, F-91405}
 \affiliation{%
 Universit\'e Paris-Sud 11, Orsay, France, F-91405}

\author{Karim Noui}
\email{karim.noui@lmpt.univ-tours.fr}

\affiliation{Laboratoire de Math\'ematiques et Physique Th\'eorique,
CNRS (UMR 7350),
F\'ed\'eration Denis Poisson,Universit\'e Fran\c{c}ois Rabelais de Tours
--- CNRS (UMR 7350),
F\'ed\'eration Denis Poisson,
Parc de Grandmont 37200 France}
\affiliation{Laboratoire APC -- Astroparticule et Cosmologie, Universit\'e Paris Diderot Paris 7, 75013 Paris, France}

\begin{abstract}
We construct and study Loop Quantum Cosmology (LQC) when the Barbero-Immirzi parameter takes the complex value $\gamma=\pm i$. 
We refer to this new quantum cosmology as complex Loop Quantum Cosmology.
We proceed in making an analytic continuation of the Hamiltonian constraint (with no inverse volume corrections) from real $\gamma$ to $\gamma=\pm i$
in the simple case of a flat FLRW Universe coupled to a massless scalar field with no cosmological constant. For that purpose, we first 
compute the non-local curvature operator (defined by the trace of the holonomy of the connection around a fundamental plaquette) 
evaluated in any spin $j$ representation and we find a new close formula for it.  This allows to define explicitly a one parameter family of
regularizations of the Hamiltonian constraint in LQC, parametrized by the spin $j$.  It is immediate to see that any spin $j$ regularization 
leads to a bounce scenario. Then, motivated particularly by previous results on black hole thermodynamics, we perform
the analytic continuation of the Hamiltonian constraint defined by $\gamma=\pm i$ and  $j=-1/2+is$ where $s$ is real. Even if the area spectrum is now
continuous, we show that  the so-defined complex LQC  removes also the original singularity which is replaced by a quantum bounce. 
In addition, the maximal density and the minimal volume of the Universe are obviously independent of $\gamma$.  Furthermore, the dynamics before and after the bounce
are no more symmetric, which makes a clear distinction between these two phases of the evolution of the Universe. 
\end{abstract}

\maketitle

\section{Motivations}
The discreteness of space at the Planck scale is certainly one of the most important and one of the most fascinating predictions of Loop Quantum Gravity (LQG) \cite{Ashtekarreview,Rovellibook,Thiemannbook}. 
In fact, it has been commonly accepted for a long time that the smooth differentiable structure  of space-time (inherited from general relativity) should be modified at this fundamental scale and replaced by a new type of quantum geometry.
However, LQG has been one of the first approaches (not to say the first one) of quantum gravity to propose a  framework where such a scenario happens precisely.  
Even if the discreteness of space has been  proved only at the kinematical level so far (where the quantum dynamics has not been taken into account), it has been rigorously established  from the \hbox{hypothesis} that
quantum states of geometry are one-dimensional excitations which live in the so-called Ashtekar-Lewandowski Hilbert space. The question of the quantum dynamics is still largely open and it is not  obvious whether  or not
the discreteness remain  at the physical level \cite{Biancha,Rovellianswer}. Nonetheless, Spin-Foam models, which are promising attempts towards the resolution of the dynamics (see the review \cite{Ale} and references therein),
seem to predict that the discreteness remains at the physical level. 

The Barbero-Immirzi parameter $\gamma$ \cite{Barbero,Immirzi} plays a central role in the scenario which has lead to the discreteness of space at the Planck scale. The physical importance of $\gamma$ shows up 
directly in the expression of the area spectrum because $\gamma$ enters in the formula of the area gap $\Delta\ell_p^2$  where $\ell_p$ is the Planck length and $\Delta=4\pi  \sqrt{3} \gamma $.
To have a deeper understanding of the significance of $\gamma$, it is essential to return at the beginnings of LQG. In \cite{Ashtekar complex}, Ashtekar showed that complex gravity can be formulated as a gauge theory where the
connection takes values in a chiral component of the complex Lie algebra $\mathfrak{sl}(2,\mathbb C)$. The complex theory has to be supplemented with reality conditions to make it equivalent to classical real
(first order) gravity. Not only gravity features a lot similarities with gauge theory in the Ashtekar formalism but also the constraints, emerging in 
the Hamiltonian formulation, are simple polynomials of the fundamental variables. This last property contrasts strongly with what happens in the usual ADM formulation where the constraints are highly non polynomial
functionals of the phase space variables. This new formulation raised the hope to perform for the first time 
the canonical quantization of gravity and then to open a promising road towards quantum gravity. LQG was born.
However, the fact that the gauge group is complex  makes the problem of quantizing the theory too complicated and even totally unsolved up to now.
Technically, the main difficulty relies on the highly non linearity of  the reality conditions that no ones knows how to handle at the quantum level.  The Barbero-Immirzi 
parameter was introduced to circumvent this problem because it allows to reformulate gravity as a real gauge theory whose gauge group is now $SU(2)$. The price to pay is that the Ashtekar-Barbero connection is no more a full
space-time connection \cite{Samuel,Alexandrov} and the Hamiltonian constraint is no more a simple
polynomial functionals of the fundamental variables. Therefore we loose some of the very interesting features of the original (complex) Ashtekar formulation of gravity. Nonetheless, one can start the 
quantization of the theory and construct at least the kinematical Hilbert space from the harmonic analysis on $SU(2)$. 
This is a deep improvement compared to the complex theory  where no Hilbert structure has been found so far even at the kinematical level. 
Therefore, introducing $\gamma$ allows to turn, at the kinematical level, the complex gauge group $SL(2,\mathbb C)$ into the real compact gauge group $SU(2)$. In that respect,
the Barbero-Immirzi parameter (when it is real) is intimately linked to the fact that  the gauge group is $SU(2)$, and then it is directly related to the discreteness of the spectra of the geometric operators
in LQG. The physical consequences of the presence of $\gamma$ in LQG are manifold. The parameter $\gamma$ plays a central role either in the bounce scenario of the Universe in Loop Quantum Cosmology (see the reviews \cite{Ashtekar,bojowald_lqc,ash_bojo_lew}) and in the recovering
of the Bekeinstein-Hawking black holes entropy (with the right famous $1/4$ factor) at the semi-classical limit \cite{RovelliBH,AshtekarBH,BHentropy3,BHentropy4,BHentropy6,BHentropy7,ENP}. 

Nonetheless, the importance of $\gamma$ at the quantum level strongly contrasts with the irrelevance of the same
parameter at the classical level where it disappears totally from the classical equations of motion. Some arguments have been proposed to explain these discrepancies \cite{interpretation} but none were totally
convincing. In a recent series of paper \cite{Geiller1,Geiller2,Geiller3,Jibril1,Jibril2,FGNP,BTZ}, we  have studied with details this puzzle  in the context of three dimensional gravity and black holes thermodynamics.
In both cases, we arrived at the conclusion that the Barbero-Immirzi parameter should not play any role in LQG, and everything converges to the idea that $\gamma$ should be fixed to its original value $\gamma=\pm i$.
In other words, we argued  that working with the imaginary value $\gamma = \pm i$ is the only choice (for the Barbero-Immirzi parameter) which will lead to a consistent quantum theory, accompanied by a proper (canonical or covariant) 
dynamics and possessing a correct semiclassical and continuum limit. This expectation is furthermore motivated by the geometrical significance carried by the complex (chiral or self-dual) Ashtekar
connection defined for $\gamma = \pm i$ and is supported by remarkable additional facts \cite{KMS,CMC,BN,Yasha1,Yasha2,Yasha3,Muxin,radiation,Amit}. 
While quantizing the self-dual theory remains an immensely hard problem because of  the reality conditions, 
one reasonable point of view is to see $\gamma$ as a regulator which should be sent back to the original complex value $\gamma = \pm i$ at some point. This idea was shown to work exactly in the context of three dimensional
gravity \cite{Jibril1,Jibril2}, and was applied  in the context of black hole thermodynamics to recover the Bekenstein-Hawking entropy \cite{FGNP} with its logarithmic corrections \cite{Amaury}.  We showed, in both cases, that
fixing the Barbero-Immirzi parameter to $\gamma = \pm i$ and performing at the same time the analytic continuation of the spin $j$ representations (coloring the spin-networks edges) to $j=-1/2 + is$ where $s$ is a real number
reproduce exactly the expected results, i.e. the right geometric spectra in three dimensional gravity and the Bekenstein-Hawking formula for the black hole entropy. The group theoretical interpretation of this analytic continuation
is clear and consists in replacing the unitary irreducible representations of $SU(2)$ by the principal (continuous) series of unitary representations of the non-compact gauge group $SU(1,1)$ to color the spin-networks
edges. Thus, these representations play apparently a very important role in three dimensional LQG and in black hole thermodynamics, but the question whether they are still relevant in the full theory in four dimensions 
is still widely open, and very complicated to answer. As a first step towards its resolution, it is interesting  to apply the same analytic continuation in the context of Loop Quantum Cosmology (LQC) where the dynamics
is a priori non trivial (contrary to three dimensional gravity where the theory is topological and contrary to black hole thermodynamics where we look at the system at equilibrium).  We hope
that such a model will help us understanding the quantization of the full complex theory of gravity and also finding our way towards the resolution of the reality conditions. 

The aim of this article is precisely to perform the analytic continuation  to $\gamma = \pm i$ in the context  of LQC. We consider the simplest cosmological model, namely the flat ($k=0$) FLRW cosmology coupled to a massless scalar 
field which plays, as usual in LQC, the role of a clock. We want to understand if the bounce scenario remains in the complex theory. 
To do so, we proceed in two steps. First of all, we compute the non-local curvature operator evaluated in any spin $j$ representations 
of $SU(2)$. This operator is the building block of the Hamiltonian constraint in LQC and has been computed so far mainly in the fundamental spin $1/2$ representation. 
To our knowledge, there is only one very interesting paper \cite{Kevin} where the question of considering spins different from $1/2$ in LQC was studied (to understand some quantum ambiguities in LQC).
Surprisingly, we find a new and simple closed formula for the non-local curvature operator in any finite dimensional representation. Then, we can perform  the analytic continuation of the curvature operator which consists in fixing $\gamma=\pm i$ and  
replacing the  integer $j$ by the complex number $j=-1/2 + is$ 
where $s\in \mathbb R$. We proceed exactly as we did in three dimensional LQG and black holes thermodynamics. We show that, even if the representations become continuous and then the area spectrum is now real, 
the curvature operator remains bounded after the analytic continuation, which means that the curvature is never singular during the history of the Universe. 
Therefore, the original singularity is resolved and generically replaced by a bounce. Contrary to what happens in real LQC,
the maximal density of the Universe (at the bounce) is independent of $\gamma$ (which in fact has been fixed previously the $\pm i$).
 
The paper is organized as follows. We start in Section 2 by a brief statement  of the main aspects of LQC. We recall that the non-local curvature operator is the building block in the construction
of the quantum dynamics. Usually, it is evaluated in the spin $1/2$ representation to achieve  the best possible coarse grained homogeneity. However, for our purpose (in order to perform
the analytic continuation), we need the expression
of the non-local curvature operator evaluated in any spin $j$ representations.  Fortunately, we obtain a new simple and closed formula for it. In Section 3, we study  the analytic 
continuation to $\gamma=\pm i$ and $j=-1/2 + is$ where $s\in \mathbb R$. We write the regularized Hamiltonian constraint (with no inverse volume corrections) in the complex theory. Afterwards we show,
in the effective theory framework first and also using the so-called exactly solvable LQC methods,  that it leads generically to 
a resolution of the initial singularity  
and to the emergence of a quantum bounce.  We finish with a discussion and some perspectives.

\section{LQC and Non-local curvature operator in any representation}
The goal of this section is twofold. First, following \cite{LQCreview}, we briefly summarize the basis (foundations, results and notations) of LQC in the simplest cosmological space-time, i.e. the flat ($k=0$) FLRW model with no cosmological constant ($\Lambda=0$)
minimally coupled to a massless scalar field. We will recall in particular that the non-local curvature operator is the key ingredient inherited from quantum geometry in the construction of LQC. This operator is the loop regularization 
of the classical curvature and is responsible
for the replacement of the original singularity by a bounce. In standard LQC, the non-local curvature operator is defined in the spin $1/2$ representation, namely it is given as the trace of the holonomy of the connection 
around a fundamental plaquette evaluated in the spin $1/2$ representation. The choice of the
spin $1/2$ is clearly motivated by technical reasons (it makes its computation much simpler and it makes the energy density of the massless scalar field positive) 
but not only. It is commonly argued that the evaluation in the spin $1/2$ representation enables to achieve  the best possible coarse grained 
homogeneity. Furthermore, the choice of the spin $1/2$ can be further motivated by the assumption that the quanta of space are minimally excited  and
that the universe expands (or contracts) by adding quanta (or removing quanta) rather than changing the representations on the spins\footnote{We would like to thank E. Wilson-Ewing for having raised  this point.}.
However, we will need the evaluation of the non-local curvature operator in any spin $j$ representation for defining the analytic continuation. This is what we are going to do in a second part, and we will obtain a new
simple and close formula for this evaluation.

\subsection{LQC for FLRW model with a massless scalar field: short review and notations}
LQC is a proposal for quantizing cosmological space-times based on the philosophy of LQG. As in the full theory, LQC is based on the (Ashtekar-Barbero) connection formulation of gravity, and the elementary
variables are given by holonomies along edges and fluxes across surfaces associated to arbitrary graphs immersed in the space, viewed as a topological space.  In the flat FLRW cosmology, the topology of space is either 
 $\mathbb R^3$ or $\mathbb T^3$, the 3-torus, and to construct properly the canonical quantization of such cosmologies, it is necessary to introduce a non-dynamical fiducial metric on the space $q$, together with
a fiducial cell $\cal C$. Typically, we choose the cell to be a cube whose length is $\ell_0$ with respect to the fiducial metric. Concerning the fiducial metric, we fix it for simplicity to the flat  metric  
${q}_a^i=\delta_a^i$ even if this is not necessary a priori. A complete discussion on the fiducial dependence of the LQC phase space structure can be found in \cite{LQCreview}. 

\subsubsection{Connection dynamics: the classical phase space}
The homogeneity and isotropy of space-time imply that we can choose the Ashtekar-Barbero connection $A_a^i$ and its conjugate electric field $E^a_i$ of the form
\begin{eqnarray}\label{def of c and p}
A_a^i=\ell_0^{-1} \, c \,  \delta_a^i \;\;\;\; \text{and} \;\;\;\; E^a_i=\ell_0^{-2} \, p \, \delta^a_i
\end{eqnarray}
where $c$ and $p$ are functions of time only.  To these geometrical variables, one adds the pair of conjugate variables $(\Phi,\Pi)$ to parametrize completely the phase space of an FLRW cosmology minimally coupled to 
a massless scalar field $\Phi$. The Hamiltonian analysis of the action of gravity coupled to the scalar field $\Phi$ leads immediately to the following definition of the Poisson brackets
\begin{eqnarray}
\{c,p\}=\frac{8\pi G \gamma}{3} \;\;\;\; \text{and} \;\;\;\; \{\Phi,\Pi\}=1\,.
\end{eqnarray}
Note that the choice of the normalization factors (depending on $\ell_0$) in (\ref{def of c and p}) makes the Poisson bracket between the dynamical variables $c$ and $p$
independent of the fiducial structure. With this choice, the quantities $c$ and $p$ are not sensitive to  the fiducial structure (neither of the fiducial metric nor of the fiducial cell).
Therefore we can fix $\ell_0=1$ without loss of generality, which is what we do from now on. Note however that  the fiducial structure does not disappear in  the quantum theory when one
considers inverse volume corrections in the Hamiltonian constraints, and the fiducial metric may also enter in the quantum uncertainties relations \cite{Ed1,Ed2}.
To complete the description of the physical phase space, we add that the lapse $N$ is a non-dynamical variable in the theory, it plays the role of a Lagrange multiplier which enforces the 
Hamiltonian constraint $C_H$ given by (note that in all this paper we will not consider the inverse volume corrections)
\begin{eqnarray}\label{localHamil}
C_H = C_{grav} +  C_{m} \;\;\; \text{with} \;\;\; C_m = \frac{ \Pi^2}{2 \vert p \vert^{3/2}} \;\;\; \text{and} \;\;\;
C_{grav}= - \frac{1}{16\pi G \gamma^2} \epsilon^{ij}{}_k \,\delta^a_i \, \delta^b_j \, \vert p \vert^{1/2} \, F_{ab}^k 
\end{eqnarray}
where the curvature of the connection $F_{ab}^k$ reduces in the FLRW model  to the simple form
\begin{eqnarray}\label{curvature}
F_{ab}^i = \epsilon_{ab}{}^i c^2 .
\end{eqnarray}
This leads immediately to the following final expression for the total Hamiltonian $H$:
\begin{eqnarray}\label{totalH}
H = NC_H \;\;\; \text{where} \;\;\; C_H=\frac{1}{2\vert p \vert^{3/2}} \left( \Pi^2- \frac{3}{4\pi G \gamma^2} p^2 c^2\right).
\end{eqnarray}
As in pure gravity, $C_H$ is a first class constraint\footnote{As  there is only one constraint in the theory, the constraint is necessarily first class.} 
which is responsible for the invariance of the under time reparametrization of the theory. Furthermore, the Hamiltonian
$H$ vanishes on shell and allows to define, as a Hamiltonian vector field, time derivatives of any function $\varphi$ of the phase space according to
\begin{eqnarray}
\dot \varphi = \{ \varphi,H \} \simeq N \{ \varphi ,C_H\}
\end{eqnarray}
where the symbol $\simeq$ refers to a weak identity (i.e. the identity holds only on the constraints surface). The freedom in the choice of the lapse function $N$ corresponds to the gauge
freedom to choose the time variable for the dynamics. 

\subsubsection{Quantization of the holonomy-flux algebra in LQC}
The quantization of such a theory (where the matter is a massless scalar field) leads generically to a singularity at the origin of the Universe  in standard Wheeler-DeWitt quantum cosmology.
The  singular behavior of usual quantum cosmologies originates from the fact that the curvature (\ref{curvature}) is unbounded from above, and then it can be infinite. This happens
precisely at the origin of the Universe. Some scenario to resolve the singularity were proposed but none are totally convincing (see \cite{LQCreview} and references therein). 
Most of them consisted in considering particular types of matter fields coupled to gravity. LQC proposes a much deeper scenario to avoid the singularity based on the fundamental structure
of space inherited from LQG. Physically, the resolution of the singularity can be viewed as a consequence of the discreteness of the geometric operators and more precisely of the existence
of a minimal spacial surface, hence a minimal spacial volume for the Universe.  In that sense, the minimal volume (or equivalently the area gap) plays the role of a UV cut-off in quantum cosmology. 
More rigorously, it is the particular structure of the kinematical Hilbert space of LQG which is deeply responsible for the resolution of the original singularity. In full LQG, the algebra of kinematical
operators form the so-called holonomy-flux algebra which admits an unique unitary irreducible representation which behaves correctly under the action of space diffeomorphims (the action is unitary) \cite{LOST}. 
Elements of the corresponding Hilbert space are one-dimensional excitations associated to a graph $\Gamma$ immersed in the space and  are
constructed from the holonomy of the connection along the edges of the graph $\Gamma$. The spin-network states form a dense basis
of this Hilbert space. They are totally  characterized by a color graph, i.e. a graph whose edges are colored with spin $j$ representations of $SU(2)$ and whose nodes are colored
with intertwiners between $SU(2)$ representations. As we will illustrate later on in the case of cosmology, the fact that fundamental excitations of quantum gravity are holonomies along edges of the connection and
not the connection itself  is intimately linked to the resolution of the singularity.

The hypothesis of homogeneity and isotropy simplify drastically the holonomy-flux algebra. Indeed, spin-network states corresponding to cosmological quantum states
are associated to graphs whose edges are parallel to the edges of the cell $\cal C$ defined above in the introduction of this section.  
Furthermore, it is commonly assumed that the edges are colored by a spin $1/2$ representation in order to achieve the best possible coarse grained homogeneity. As a consequence, holonomies $h_\ell$ of the
(homogeneous and isotropic) Ashtekar-Barbero connection (\ref{def of c and p}) along a line $\ell$ parallel to one of the edges of the cell $\cal C$ and evaluated in the spin $1/2$ representation
are the building blocks of kinematical states in LQC. If we denote by $\mu$ the length of $\ell$ with respect to the fiducial metric $q$, it is immediate to show that
\begin{eqnarray}
h_\ell = \exp (\int_\ell dx^a \, A_a^i \, \tau_i)= \exp \left( \mu  \, c \, \tau_\ell \right) = \cos(\frac{\mu c}{2}) + 2 \sin (\frac{\mu c}{2}) \tau_\ell
\end{eqnarray}
where $\ell$ has been identified with its direction and then takes value in the set $\{1,2,3\}$. Here $\tau_\ell$ are the generators of $\su(2)$ evaluated in the fundamental
representation, they satisfy  the Lie algebra relation $[\tau_i,\tau_j]=\epsilon_{ij}{}^k\tau_k$ and the normalization property $\tau_\ell^2=-1/4$ for any $\ell$.  
We can choose a basis such that they are explicitly given by the following $2\times 2$ matrices:
\begin{eqnarray}
\tau_1=\frac{1}{2} \left( \begin{array}{cc} i & 0 \\ 0 & -i \end{array} \right) \;\;\;,\;\;\;
\tau_2=\frac{1}{2} \left( \begin{array}{cc} 0 & 1 \\ -1 & 0 \end{array} \right) \;\;\;,\;\;\;
\tau_3=\frac{1}{2} \left( \begin{array}{cc} 0 & i \\ i & 0 \end{array} \right) \;.
\end{eqnarray}
As we have just said in the general discussion above, fundamental variables in LQC kinematics are 
the elements $h_\ell$, or equivalently they are (complex) 
exponentials of the connection $c$ and not the connection itself. It is important to note that one cannot recover,  for instance, the connection by a derivation of $h_\ell$ with respect to $\mu$ because the 
holonomy-flux algebra representation is  not weakly continuous, hence deriving the holonomy operator $h_\ell$ is not a well-defined action. 
For that reason, the Hamiltonian constraint $C_H$, as it is written in (\ref{totalH}), is not a well-defined operator  acting on the LQC kinematical states 
and  needs to be regularized. The regularization is constructed  as in gauge theory where the curvature (\ref{curvature}) is expressed in terms of holonomy variables $h_\ell$ only. 

\subsubsection{Regularization of the curvature operator}
The key point to find a suitable regularization of the curvature (\ref{curvature}) is the formulae
\begin{eqnarray}\label{Ftracej}
F_{ab}^\ell = \frac{1}{\tr_j(\tau_1^2)} \lim_{\text{Ar}_\Box \rightarrow 0} \frac{\tr_j \left( h_{\Box_{ab}} \tau^\ell \right)}{\text{Ar}_\Box} \;\;\; \text{with} \;\;\; \tr_j(\tau_1^2)= -\frac{j(j+1)(2j+1)}{3}=-\frac{d(d^2-1)}{12}
\end{eqnarray}
where $h_{\Box_{ab}}$ is the holonomy around a square plaquette whose edges are parallel to the directions $a$ and $b$ of the fiducial $\cal C$, $\text{Ar}_\Box$ its area with respect to the flat fiducial metric $q$
and $\tr_j(X)$ denotes the trace of the element $X$ (which belongs to the enveloping algebra of $\su(2)$) in the spin $j$ representation of dimension $d=2j+1$. The holonomy around the plaquette can be obviously expressed in terms of the holonomies $h_a(\bar{\mu})$
and $h_b(\bar\mu)$ along the edges of length $\bar{\mu}$ (with respect to the fiducial metric fixed such that $\ell_0=1$ from the beginning) and directions $a$ and $b$ respectively as follows:
\begin{eqnarray}\label{hBox}
 h_{\Box_{ab}} = h_a(\bar{\mu}) h_b(\bar{\mu}) h_a(\bar{\mu})^{-1} h_b(\bar{\mu})^{-1}.
\end{eqnarray}
From the quantum point of view, the limit where $\text{Ar}_\Box$ vanishes is not defined, again because of the lack of weak continuity property of the kinematical Hilbert space representation (known as the 
Ashtekar-Lewandowski representation). Physically, we can say that the limit does not exist because  the existence of a minimal area in the theory or equivalently an area gap forbids us to send the area of any
surface to zero. At most, we can send the area to its minimal value (the discreteness of the area spectrum is closely related to the continuity properties of the Ashtekar-Lewandowski representation).   
Determining the minimal value of the area $\text{Ar}_\Box$ is equivalent to determining the expression of the minimal length $\bar{\mu}$ of its edges. 
As we see  from the notations, we work in the so-called improved $\bar{\mu}$-scheme where it was shown that \cite{improvedLQC}
\begin{eqnarray}
\bar{\mu} = \ell_p \, \frac{\Delta^{1/2}  }{ \vert p \vert^{1/2}}
\end{eqnarray}
$\Delta$ being the quantum of area (in Planck unit with $\ell_p^2=G$) carried by each edge of the spin-network state representing a homogeneous and isotropic quantum state of geometry. 
The consequence is that the classical curvature is replaced in LQC by a non-local curvature operator which clearly needs ordering prescriptions to be well-defined.

To be more explicit, let us fix the spin in (\ref{Ftracej}) to the value $j=1/2$ as it is usually done in LQC. This choice is motivated by the fact the quantum states in quantum cosmological  are colored with 
spin $1/2$ representations (to achieve the best possible coarse grained geometry), and then it is natural (and technically simpler) to define the curvature operator in term of holonomies evaluated in the 2-dimensional representation as well. Nonetheless, we will shortly 
discuss this choice in the following subsection. When $j=1/2$, $\Delta$ reduces to the area gap $\Delta=4\pi  \sqrt{3} \gamma$  and the expression of the  holonomy (\ref{hBox}) simplifies as follows:
\begin{eqnarray}\label{hboxab}
h_{\Box_{ab}} & = & \cos({\bar{\mu} c})  + \frac{1}{2} \sin^2({\bar{\mu} c}) 
+   \sin^2({\bar{\mu} c})  \epsilon_{ab}{}^{c}\, \tau_c +  (1-\cos({\bar{\mu} c}))  \sin({\bar{\mu} c}) (\tau_a - \tau_b) 
\end{eqnarray}
The calculation of the non-local curvature operator $\hat{F}_{ab}^\ell$ (the hat notation allows to distinguish the  local curvature from the quantum
non-local one) follows immediately:
\begin{eqnarray}\label{nonlocalF}
\hat{F}_{ab}^\ell =   \frac{\sin^2(\bar{\mu}c)}{\bar{\mu}{}^2} \, \epsilon_{ab}{}^\ell + \frac{(1-\cos({\bar{\mu} c}))  \sin({\bar{\mu} c}) }{\bar{\mu}{}^2}  (\delta_{a}^\ell - \delta_b^\ell).
\end{eqnarray}
The second term in the r.h.s. of the previous equation is often omitted. The reason is that the curvature appears in the Hamiltonian constraint (\ref{localHamil}) in the form 
$\epsilon^{ab}{}_\ell F_{ab}^\ell$ and then this second term disappears from the Hamiltonian because of symmetry properties. In fact, it is  implicitly assumed in the literature that $\ell$ is different from $a$ and
$b$ when one writes the components $F_{ab}^\ell$ of the curvature. 

To simplify  the study of the quantum dynamics, it is convenient to change variables and to consider  $b=c/\vert p \vert^{1/2}$ instead of the original connection $c$. The reason is that, at the classical level, 
$b$  is  conjugated to the volume $V=\vert p \vert^{3/2}$ in the sense that $\{b,V\}=4\pi \gamma G$, and this simplifies highly
the resolution of the Hamiltonian constraint as we are going to see shortly. When expressed in terms of $b$ and $V$, the total Hamiltonian (\ref{totalH}) becomes 
\begin{eqnarray}\label{nonlocalH}
\hat{H} = N \left( H_m  - \frac{3}{8\pi G \gamma^2}  \frac{\sin^2(\lambda b)}{\lambda^2} V \right) \;\;\; \text{with} \;\;\; H_m= \frac{\Pi^2}{2 V} \;\;\text{and}\;\;\;
\lambda =\ell_p \sqrt{\Delta}.
\end{eqnarray} 
This expression is the starting point of the study of the dynamics in LQC. There remain  ordering ambiguities to understand when we promote the Hamiltonian constraint as
an operator acting on a suitable Hilbert space. The reason is that the quantum operators $\hat{b}$ and $\hat{V}$ corresponding respectively to the quantizations of $b$ and $V$ are not commuting. These aspects have been
deeply studied and were nicely reported in \cite{robustness} and \cite{LQCreview} for instance.

\subsubsection{Effective dynamics}
\label{effective}
There exists a simple way to see how the Hamiltonian constraint (\ref{nonlocalH}) leads generically to a bouncing Universe  in LQC without entering into the details of the quantization.
The method we are talking about is known as the effective analysis of the dynamics.  However, even if this method is qualitatively very interesting, it misses some important aspects that can be
analyzed only from a rigorous analysis of the quantum theory. 

In the effective analysis, one interprets the Hamiltonian $\hat{H}$ as a classical function  and not as a quantum operator. Then
we immediately get the energy density of the scalar field from $\hat{H}=0$:
\begin{eqnarray}\label{densityrho}
\rho = \frac{H_m}{V}=\frac{3}{8\pi G \gamma^2}  \frac{\sin^2(\lambda b)}{\lambda^2} .
\end{eqnarray}
Therefore, we obtain directly a first indication of the existence of a bounce because $\rho$, viewed as a function of $b$, is now bounded (compared to the standard Wheeler-DeWitt quantum cosmology)
with a maximum  $\rho_{max}=3/(8\pi G \gamma^2 \lambda^2)$.
This is fundamentally due to the finiteness of $\lambda$ and then to the existence of an area gap in LQG. If the density energy is bounded, then there is no more singularity and 
they may exist a minimal volume of the Universe. In fact, we can show that this is generically  the case from a careful analysis of the quantum dynamics (under some conditions on the quantum states).

There is a second (certainly more spectacular) indication  for the resolution of the initial singularity and the emergence of a bounce scenario. It is based on the
the analysis of the modified Friedmann equation induced by the dynamics of LQC. The modified Friedmann equation is very easy to obtain and relies on the calculation
of the time derivative of the volume $\dot{V}= \{V,C_H\}$ where the time parameter is the cosmic time (i.e. $N=1$). From this straightforward computation, we obtain (the details can be found in \cite{Corichi} for instance)
\begin{eqnarray}
\left( \frac{\dot{a}}{a}\right)^2 = \left( \frac{\dot{V}}{3V}\right)^2= \frac{8\pi G}{3} \rho \left( 1 - \frac{\rho}{\rho_{max}}\right).
\end{eqnarray}
The scale factor $a$ is related to the volume by $V=a^3$ and $(\dot{a}/a)$ represents the usual Hubble parameter. The modified Friedmann equation has been studied intensively \cite{effective} and it can be shown precisely that
it describes  a bouncing Universe. Without going to the details, it is easy to see that the Universe admits a minimal volume which is reached when $\rho=\rho_{max}$ (whence $\dot{V}=0$). Furthermore,
everything happens as if the quantum gravity effects manifest themselves at the origin as a pressure which forbids the Universe from collapsing and then removes the original singularity.

\subsection{The non-local curvature operator evaluated in any spin $j$ representation}
The regularization of the curvature operator is the cornerstone of LQC. It is mathematically based on the techniques used in full LQG and it is 
physically responsible for the resolution of the initial singularity. However, the regularization procedure is not unique and suffers from ambiguities. The main ambiguity relies on
the choice of the spin $j$ representation in (\ref{Ftracej}). By construction, the classical curvature does not depend on that choice  but the value of the spin $j$ becomes relevant at the quantum level,
when the limit $\mu \rightarrow 0$ is replaced by $\mu \rightarrow \bar{\mu}$ in (\ref{Ftracej}). Some aspects of this ambiguity were studied in \cite{Kevin} and  similar ones in the full theory were discussed in \cite{Aleambiguity}.
The article \cite{Kevin} has been, to our knowledge, the only one to discuss the choice of the spin in the regularization of the curvature operator for LQC. 

To have a deeper understanding of the role of $j$ in the regularization of the curvature, it would be much better to have an explicit and simple expression of the trace $\tr_j \left( h_{\Box_{ab}} \tau^\ell \right)$ involved
in the definition of the non-local curvature operator, which is missing up to our knowledge (in \cite{Kevin} the trace was expressed as  a finite sum only).  
But we are going to provide in this section a new and simple expression for the trace in any spin $j$ representation. This new formula is clearly a good starting point for the study of the ambiguity we mentioned above
 and its effects on the dynamics of LQC. However, we are not interested in studying this aspect here that we postpone for later \cite{Ed}. We are more interested in the analytic continuation of 
 LQC to complex Barbero-Immirzi parameters, more precisely to $\gamma=\pm i$. However, we argued in the introduction that fixing $\gamma = \pm i$ in LQG is meaningless a part if we send at the same time the discrete spin $j$
 representations (coloring the spin-network edges) to the complex values $j=-1/2 + is$. To adapt this construction in the context of LQC, we need to write the non-local curvature, hence $\tr_j \left( h_{\Box_{ab}} \tau^\ell \right)$, 
 as an analytic function of $j$ (and also $\gamma$). This is exactly what we are going to do in this section.
 
 We proceed as follows. To lighten the notations,  we choose to compute $\tr_j \left( h_{\Box_{12}} \tau_3 \right)$, the other components will be directly obtained from permutations between the indices. First we remark
 that
 \begin{eqnarray}\label{defh123}
 \tr_j \left( h_{\Box_{12}} \tau_3 \right) = \frac{\partial \,  \tr_j \left(h_{123}(\varepsilon) \right)}{\partial \varepsilon}\vert_{\varepsilon=0}\;\;\; \text{where} \;\;\; h_{123}(\varepsilon)= 
 h_{\Box_{12}} g_3(\varepsilon) \;\; \text{and} \;\; g_3(\varepsilon)=\exp(\varepsilon \tau_3).
 \end{eqnarray}
Then the calculation reduces in computing the trace of the group element $h_{123}(\varepsilon) \in SU(2)$ in the spin $j$ representation. We know from representation theory
that such a trace is given by the  character 
\begin{eqnarray}
\tr_j \left( h_{123}(\varepsilon)  \right) = \chi_j\left(\theta(\varepsilon)\right) = \frac{\sin \left(d\theta(\varepsilon)\right)}{\sin \left(\theta(\varepsilon)\right)}
\end{eqnarray}
where $\theta(\varepsilon) \in [0,\pi]$ is the conjugacy class of $h_{123}(\varepsilon)$. 
Now, the point is that we can obtain the angle $\theta(\varepsilon)$  using the $\su(2)$ fundamental  representation where  $h_{123}(\varepsilon)$ is a 2-dimensional matrix which can be easily computed from
(\ref{hboxab}).  Then a  direct calculation leads to:
\begin{eqnarray}\label{thetab}
\chi_{1/2}\left(\theta(\varepsilon)\right) & = & 2 \cos\left(\theta(\varepsilon)\right) = \tr_{1/2} (h_{123}(\varepsilon)) \nonumber \\
 & = & (2 \cos(\lambda b) + \sin^2(\lambda b)) \cos(\frac{\varepsilon}{2}) - \sin(\frac{\varepsilon}{2}) \sin^2(\lambda b) .
\end{eqnarray}
The expression of (\ref{defh123}) in terms of the variable $b$  follows immediately:
\begin{eqnarray}
\tr_j \left( h_{\Box_{12}} \tau_3 \right)   =  \frac{\partial}{\partial \theta} \left( \frac{\sin(d\theta)}{\sin \theta}\right) \times \left(\frac{\partial \theta}{\partial \varepsilon}\right)(\varepsilon=0) 
  =   \frac{\sin^2(\lambda b)}{4 \sin \theta} \times \frac{\partial}{\partial \theta} \left( \frac{\sin(d\theta)}{\sin \theta}\right) 
\end{eqnarray}
Note that the angle $\theta$ with no $\varepsilon$ dependence is defined by $\theta=\theta(0)$. Its expression in terms of $b$ is obtained from the relation (\ref{thetab}) between $b$ and $\theta(\varepsilon)$
with $\varepsilon=0$, and a short calculation leads to:
\begin{eqnarray}\label{relation theta b}
\sin^2\left( \frac{\theta}{2}\right) = \sin^4 \left( \frac{\lambda b}{2}\right)
\end{eqnarray}
which defines totally $\theta \in [0,\pi]$ in terms of $b$.  As an immediate consequence, the relevant components of the non-local curvature operator 
regularized using the spin $j$ representation (of dimension $d$) are given by
$\hat{F}_{ab}^\ell=\epsilon_{ab}{}^\ell \hat{F}_{12}^3$ (when $\epsilon_{ab\ell}\neq 0$) with:
\begin{eqnarray}
\hat{F}_{12}^3 = - \frac{3}{d(d^2-1)}  {\vert p \vert}\frac{\sin^2(\lambda b)}{\lambda^2} \; \frac{1}{ \sin \theta} \times  \frac{\partial}{\partial \theta} \left( \frac{\sin(d\theta)}{\sin \theta}\right) 
\end{eqnarray}
Note that  $\lambda=\ell_p \sqrt{\Delta}$ should take different values depending on the choice of the spin $j$. Indeed, in the $\bar{\mu}$-scheme, it has been argued that 
$\Delta$ represents the quantum of area (in Planck units) carried  by spin-network describing LQC quantum states. If this is indeed the case, then $\Delta=4\pi\gamma \sqrt{d^2-1}$. 
One could also argue that $\Delta$ is simply the area gap (the minimal area) in Planck units of LQG in which case it is independent of $j$.
 
 \medskip 
 
\begin{figure}
\begin{center}
	\includegraphics[scale=0.4]{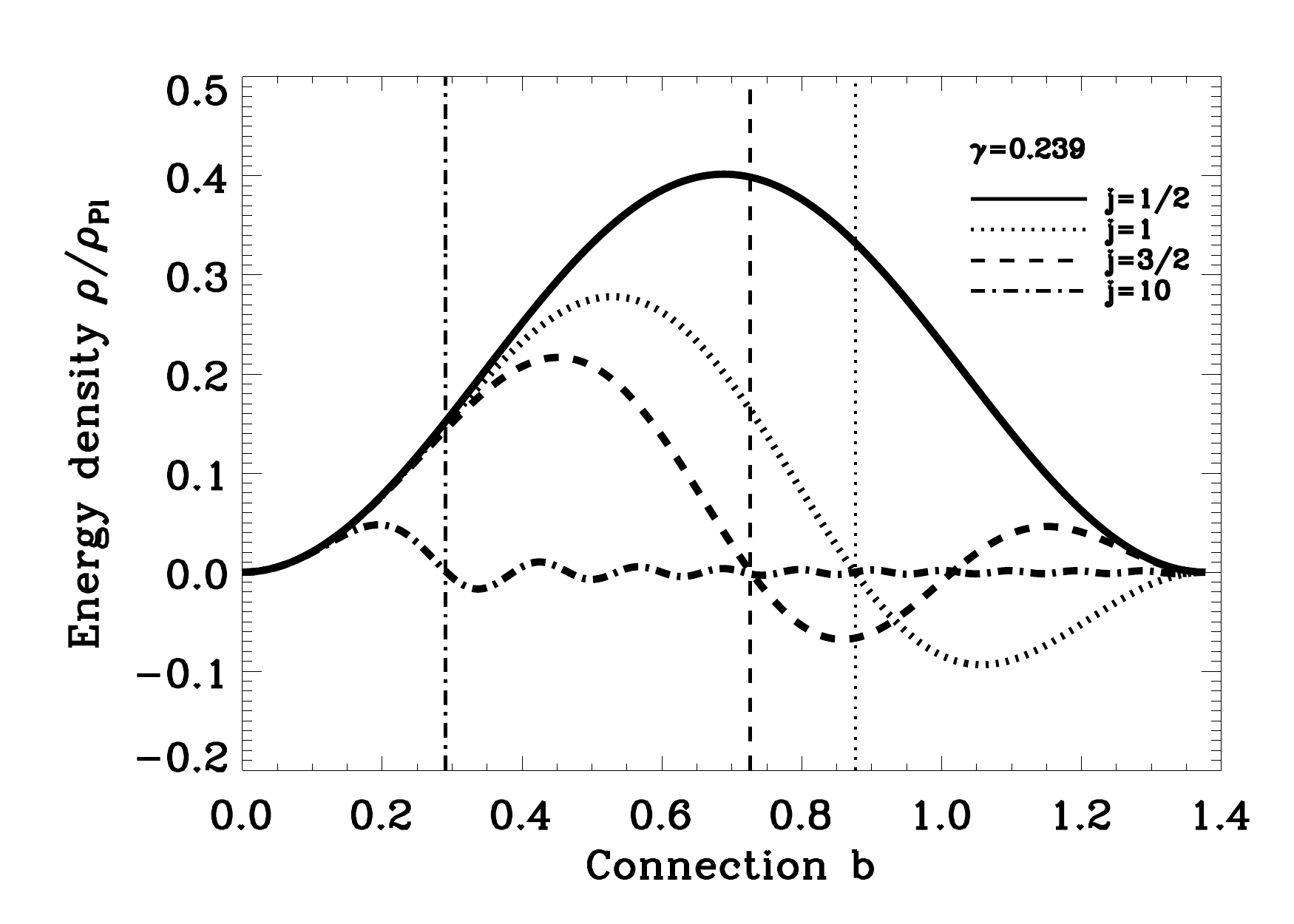}\includegraphics[scale=0.4]{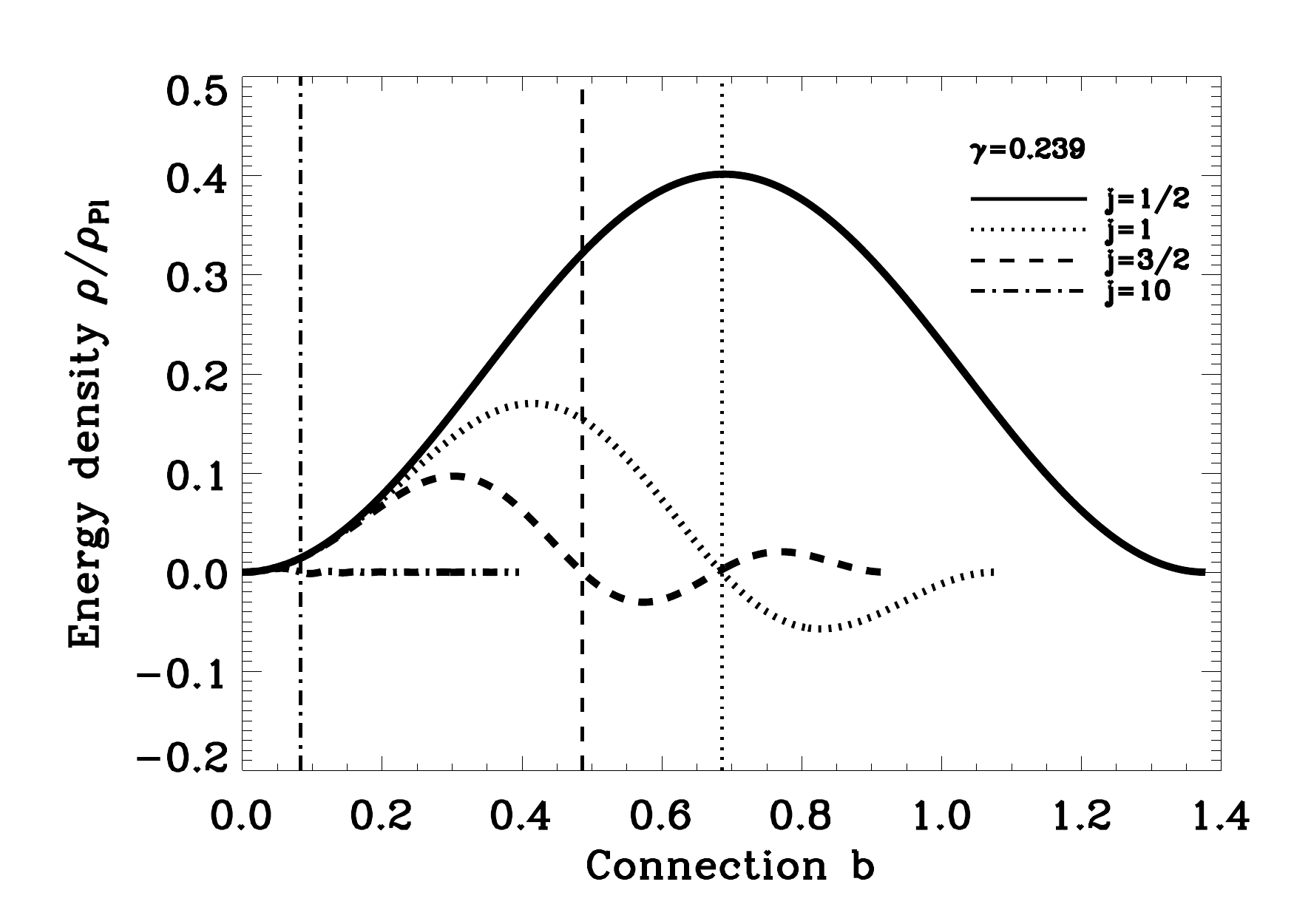}
	\caption{Energy density as a function of $b$ as derived from the vanishing of the Hamiltonian constraint, $C_H=0$, in Planck units $\rho_\mathrm{Pl}=1/G^{2}$. In the left panel, $\Delta$ is assumed to be given by the area gap, $\Delta=4\pi \gamma \sqrt{3}$, and therefore does not depend on the chosen representation $j$. In the right panel, $\Delta$ is given by the quantum of area associated to the chosen representation $j$, {\it i.e.} $\Delta=4\pi \gamma \sqrt{d^2-1}$. This second option makes the maximal energy density lower. This also makes the range of values of $b$ to depend on $j$ since $0\leq b\leq\pi/\lambda(j)$ with $\lambda(j)=\ell_p(4\pi\gamma\sqrt{d^2-1})^{1/2}$. For both cases, the dotted, dashed, and dashed-dotted vertical lines flag the value of $b$ for which the energy density first becomes negative for $j=1,~3/2$ and 10, respectively.}
	\label{fig:real}
\end{center}
\end{figure}

 We finish this section with some examples. 
 \begin{enumerate}
 \item When $j=1/2$ (d=2) we recover immediately the expression (\ref{nonlocalF}).
 \item When $j=1$ (d=3) we find after a straightforward calculation:
 \begin{eqnarray}
 \hat{F}_{12}^3 = \vert p \vert  \, \frac{\sin^2(\lambda b)}{\lambda^2}  \cos \theta =  \vert p \vert  \, \frac{\sin^2(\lambda b)}{\lambda^2} \left( 1 - 2 \sin^4 \left( \frac{\lambda b}{2}\right) \right).
  \end{eqnarray}
 \end{enumerate}
 For any finite values of $j$, the dynamics of LQC leads to a resolution of the singularity since the non-local curvature is bounded from above, and so is the energy density. However, the dynamics is modified and depends strongly on the choice of the representation. These aspects have been explored in \cite{Kevin} and are revisited in \cite{Ed}. Here we briefly underline some difficulties which may arise for $j>1/2$. From $C_H=0$ and defining the energy density of the matter content by $\rho=H_m/V$ one easily gets the modified Friedmann equation relating $\rho$ to the non-local curvature for any $j$:
 \begin{equation}
 	\rho=\frac{9}{8\pi G\gamma^2d(d^2-1)} \frac{\sin^2(\lambda b)}{\lambda^2} \; \frac{1}{ \sin \theta} \times  \frac{\partial}{\partial \theta} \left( \frac{\sin(d\theta)}{\sin \theta}\right).
\end{equation}
The shape of the r.h.s. of the above is displayed in Fig. \ref{fig:real} for four values of the spin representation, $j=1/2,~1,~3/2$ and 10. This is shown in Planck units defining $\rho_\mathrm{Pl}=1/G^2$. The left panel is obtained by choosing $\Delta$ from the minimal area, $\Delta=4\pi \gamma \sqrt{3}$, and the right panel is obtained by choosing $\Delta$ from the quantum of area, $\Delta=4\pi  \gamma \sqrt{d^2-1}$. For any spin, the non-local curvature is bounded from above, solving for the singularity, but it also admits negative values for $j>1/2$. This would mean that the energy density itself could become negative-valued. However, from its definition, one easily figures out that $\rho=H_m/V=\Pi^2/(2V^2)$ which is positive-valued. 
Then, the variable $b$ should be restricted to belong to some range for which the energy density is positive. This still raises some issues since the Hubble parameter is non-vanishing at the smallest value of $b$ for which the energy density would become negative, called $b_s$ hereafter. (Those values for $j>1/2$ are depicted by vertical lines in Fig. \ref{fig:real}.) The Hubble parameter is indeed given by $(\dot{a}/a)\propto\frac{\partial\rho}{\partial b}$ which is non-zero when the sign of $\rho$ firstly changes at $b_s$. This is clearly shown for {\it e.g.} $j=1$ in Fig. \ref{fig:real} where the change of sign occurs at $b_s\sim0.88$. This means that there is a priori no dynamical reason for this change of sign not to be crossed. (We stress that the point $b_s$ would be only asymptotically reached, and therefore never crossed, if the Hubble parameter were vanishing at $b_s$.)
 
\section{Complex Loop Quantum Cosmology}
Now we have all the necessary tools to perform the analytic continuation and to study the dynamics of LQC when  $\gamma=\pm i$. In a first subsection, we will
construct the analytic continuation of the non-local curvature operator. In a second subsection, we will study the induced effective theory.
We see immediately from the previous section that if we set $\gamma=\pm i$ without changing the representations $j$, the curvature is no more bounded and the original singularity 
is not removed. If  we add the condition  $j=-1/2 + is$ with $s$ real, not only the area spectrum remains positive but also we recover the bouncing scenario but the precise History of the Universe is very different 
 from real LQC. For instance, the energy density is no more periodic which implies that the History of the Universe is not periodic neither as it is in real LQC.

The case $s=0$ (equivalently $j=-1/2$)
is the analogous of $j=1/2$ in the sense that it allows to achieve the best possible coarse grained  geometry in the complex theory. Therefore, we study more precisely this situation and shows that, exactly as in usual LQC,
the energy density is always positive. The dynamics before and after the bounce are totally different which makes a clear distinction between the contracting and the expanding phases of the evolution
of the Universe. This contrasts with what happens in usual LQC where the dynamics after and before the bounce are totally symmetric. With that respect, the complex theory offers a more satisfying scenario.

\subsection{Analytic continuation: $\gamma=\pm i$ and $j=-1/2 + is$}
We proceed now to the analytic continuation. To do so, we need to make explicit the $\gamma$-dependence in the Hamiltonian constraint.   Indeed, $\gamma$ is hidden in the connection variable
$c$ and appears clearly when we recall its relation to the usual scalar factor $a$ which is  $c=\gamma \dot{a}$. As a consequence, the variable $b$ is also proportional to $\gamma$ and then it will be useful 
in this section to make this proportionality relation explicit. For that reason, from now on, we change slightly the notations as follows:
\begin{eqnarray}
c \longrightarrow c=\gamma \tilde{c} \;\;\; \Longleftrightarrow b \longrightarrow b=\gamma \tilde{b} \,.
\end{eqnarray}
In fact, $\tilde{c}$ represents the extrinsic curvature in homogeneous and isotropic geometries.
When we fix the Barbero-Immirzi parameter to $\gamma = \pm i$, the variable $b$ becomes purely imaginary.  
If we naively believe this is the only changing in the theory (to recover self-dual variables), then we could directly
implement it in the expression (\ref{nonlocalH}) of the regularized (with spin $1/2$) Hamiltonian constraint and we would see 
immediately that the energy density (\ref{densityrho}) is no more bounded from above. The consequence is that we would loose
the bouncing scenario  and the original singularity would not be resolved in complex LQC. 

In fact, we miss an important ingredient. When $\gamma=\pm i$ in LQG, we must change at the same time the  representations $j$ coloring the spin-networks
according to  $j=-1/2 + is$ with $s$ real, for the area eigenvalues to remain real (or equivalently for the area operator to be a well-defined self-adjoint operator). Indeed,
the quantum of area $a(j)$ carried by a link colored with the spin $j$ remains real under this  change, as seen in the following
\begin{eqnarray}
a(j)=8\pi \gamma \ell_p^2 \sqrt{j(j+1)} \longrightarrow a(s)= \pm 4\pi \ell_p^2 i\sqrt{-(s^2+1/4)}=4\pi \ell_p^2 \sqrt{s^2+1/4}
\end{eqnarray} 
where we choose the square root such that the area is positive. Note that the cases $\gamma=+i$ and $\gamma=-i$ are totally equivalent and, without loss of generality, we will
consider only  the case $\gamma=+i$ in the rest of the article. An important consequence is that, even if their expressions change, $\Delta$, hence $\lambda$ (\ref{nonlocalH}), are non negative real constants. 
They may depend on $s$ exactly as they may depend on the representation $j$ in usual real LQC. But for the reasons we have already mentioned above, we take $\Delta$ to be simply the area gap
which becomes in the complex theory
\begin{eqnarray}\label{complexagap}
\Delta  = 2\pi \ell_p^2.
\end{eqnarray}
It is obviously independent of $\gamma$, in that sense it is universal. 

At this point, to obtain the complex Hamiltonian constraint, it remains to understand how $\theta$ changes when $\gamma=i$ in analyzing the relation (\ref{relation theta b}). To that aim, we first
notice that (\ref{relation theta b}) is totally equivalent to the relation (on the r.h.s.)
\begin{eqnarray}\label{identitybtheta}
\sin^2\left( \frac{\theta}{2}\right) = \sin^4 \left( \frac{\lambda b}{2}\right) \Longleftrightarrow \sin^4\left( \frac{\theta}{2}\right) = \sin^8 \left( \frac{\lambda b}{2}\right)
\end{eqnarray}
when $\theta$ and $b$ are real. However, when $b$ is imaginary, these two relations are not equivalent and lead to  inequivalent definitions of the angle $\theta$. 
Fortunately, only the second relation gives a consistent and simple definition of $\theta$ which is 
\begin{eqnarray}\label{tildetheta}
\theta \longrightarrow i \tilde{\theta} \in i \mathbb R \;\;\; \text{with} \;\;\; \sinh\left( \frac{\tilde{\theta}}{2}\right) = \sinh^2 \left( \frac{\lambda \tilde{b}}{2}\right)
\end{eqnarray}
where we assumed that $\tilde{\theta}>0$ for simplicity but without loss of generality. If we have started with the first identity in (\ref{identitybtheta}), we would have obtained a complex
angle $\theta$ (with a non zero real and imaginary parts) which would have produced a complex energy density which is of course physically irrelevant.  

The expression of the local curvature operator in the complex theory follows immediately 
\begin{eqnarray}
\hat{F}_{12}^3 \longrightarrow \hat{\tilde{F}}_{12}^3 =  \frac{3}{s(s^2+1)} \vert p \vert \frac{\sinh^2(\lambda \tilde{b})}{\lambda^2} \frac{1}{\sinh \tilde{\theta}} \times  \frac{\partial}{\partial \tilde{\theta}} 
\left( \frac{\sin(s \tilde{\theta})}{\sinh \tilde{\theta}}\right)
\end{eqnarray}
where $\tilde{\theta}>0$ has been defined above (\ref{tildetheta}). As a consequence, the regularized Hamiltonian constraint (depending on the choice of $s$) is given by
\begin{eqnarray}
H \longrightarrow \tilde{H}=N \left( H_m + \frac{3}{8\pi G} \vert p \vert^{1/2} F_{12}^3 \right)
\end{eqnarray}
where the matter field energy $H_m=\Pi^2/2V$ is unchanged. We supplement the expression of the Hamiltonian constraint with the Poisson brackets between the new (complex) geometric variables (in fact only the
connection variable is new, the electric field is unchanged)
\begin{eqnarray}
\{ \tilde{c},p\}=\frac{8\pi G}{3} \Longleftrightarrow \{\tilde{b},V\}=4\pi G
\end{eqnarray}
and we obtain the complete definition of the dynamics of LQC in complex Ashtekar variables. The Poisson brackets involving the scalar field degrees of freedom are unchanged.

\subsection{Effective dynamics in the complex theory}
This section aims at studying the effective dynamics induced by the Hamiltonian we have just described in the previous part. We follow exactly the classical analysis recalled
in subsection (\ref{effective}). First, we compute the energy density $\tilde{\rho}$ of the scalar field in the complex theory:
\begin{eqnarray}
\tilde{\rho} = \frac{H_m}{V} & =  &-\frac{9}{8\pi G s(s^2+1)} \frac{\sinh^2(\lambda \tilde{b})}{\lambda^2}  \frac{1}{\sinh \tilde{\theta}} \times  \frac{\partial}{\partial \tilde{\theta}}  
\left( \frac{\sin(s \tilde{\theta})}{\sinh \tilde{\theta}}\right). \nonumber \\
 & = & \frac{9}{8\pi G s(s^2+1)} \frac{\sinh^2(\lambda \tilde{b})}{\lambda^2} \frac{\coth \tilde{\theta} \sin(s \tilde{\theta}) - s \cos(s\tilde{\theta})}{\sinh^2 \tilde{\theta}}.
\end{eqnarray}
Here we have not specified yet the value of the representation $s$. We will discuss  this particular point later on. Before going further, we remark that the energy density  
$\tilde{\rho}$ is much simpler when it is expressed in terms of $\tilde{\theta}$ than in terms of the complex (modified)  connection $\tilde{b}$. Physically, $\tilde{\theta}$ represents
somehow the curvature because it is the conjugacy class of the holonomy $h_{ab}$ around plaquettes. In that sense, the curvature might be a more natural variable than the connection
itself to study the dynamics. Note that this remark holds also in the usual real case when the non-local curvature operator is regularized with an arbitrary spin $j$.

\begin{figure}
\begin{center}
	\includegraphics[scale=0.4]{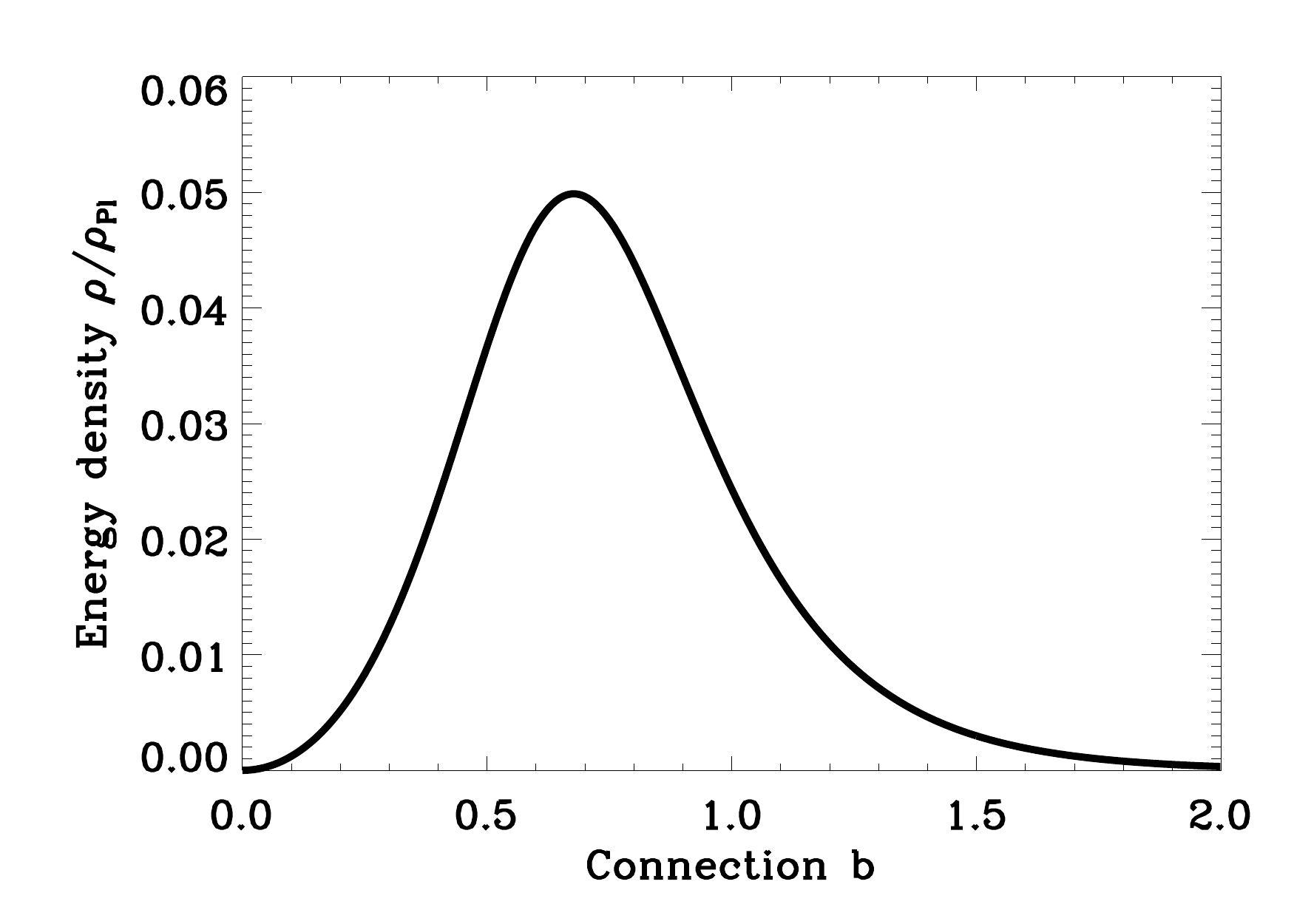}\includegraphics[scale=0.4]{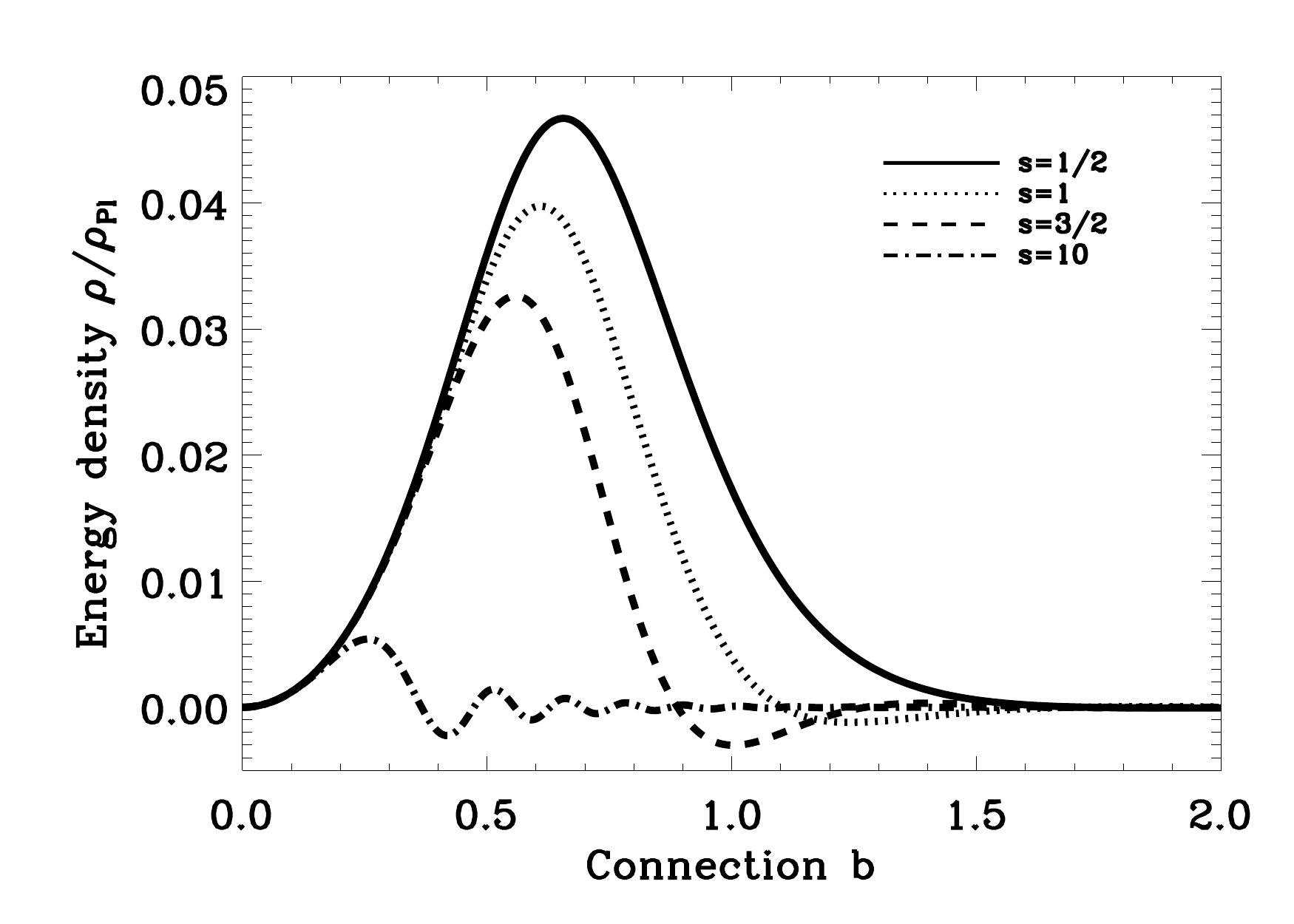}
	\caption{Energy density as a function of $\tilde{b}$ for complex loop quantum cosmology for the lowest representation $s=0$ (left panel) and for $s=1/2,~1,~3/2$ and 10 (right panel).}
	\label{fig:complex}
\end{center}
\end{figure}

The bounce occurs because the energy density remains bounded after the complexification procedure. To see this is indeed the case,  we show that $\tilde{\rho}$ is not singular 
at $\tilde{b}=0$ or equivalently at $\tilde{\theta}=0$  (even if there is a $\sinh^2 \tilde{\theta}$ in the denominator) and we also show that it vanishes when  $\tilde{b}$ becomes big. For that purpose, we compute the
following small $\tilde{b}$ and large $\tilde{b}$ expansion:
\begin{eqnarray}\label{asympenergy}
\tilde{\rho} & \sim & \frac{3}{8\pi G} \tilde{b}{}^2 \;\;\; \text{when} \;\;\tilde{b} \ll 1 \\
\tilde{\rho} & \sim & \frac{9}{8\pi G s(s^2+1) \lambda^2} \exp(-\tilde{\theta}) \left( \sin(s\tilde{\theta}) - s \cos(s\tilde{\theta})\right) \;\;\; \text{when} \;\;\tilde{b} \gg 1\,.
\end{eqnarray}
As $\lim_0 \tilde{\rho}=0= \lim_\infty \tilde{\rho}$, and as $\tilde{\rho}$ admits no pole, then $\tilde{\rho}$ is necessary bounded, hence the bouncing scenario. The shape of the energy density as a function of $\tilde{b}$ is depicted in Fig. \ref{fig:complex} for $s=0$ (left panel) and for selected values of $s>0$ (right panel), explicity showing that $\tilde{\rho}$ admits a maximal value. As it is already the case in real LQC, the energy density as a function of $\tilde{b}$ may 
become negative-valued for $s>0$.
However, contrary to the real case with $j=1/2$, one cannot write explicitly a modified Friedmann equation which would involve the Hubble constant and the energy
density only. It is only when $\tilde{b}\ll 1$ (equivalently when $\tilde{\theta}\ll 1$) that one recovers trivially the standard classical Friedmann equation (in cosmic time, i.e. $N=1$) 
\begin{eqnarray}
\dot{V} = \{ V, C_H\} = V \{\tilde{\rho},V\} = 3V\tilde{b} \Longrightarrow \left( \frac{\dot{a}}{a}\right)^2 =  \left( \frac{\dot{V}}{3V}\right)^2 = \frac{8\pi G}{3} \tilde{\rho}
\end{eqnarray}
as expected.  In the quantum regime, there is no such an exact equation. We can obviously say that the bounce occurs when the energy density is maximal, but
even a close formula for the maximal energy density does not exist. For these reasons, it is very complicated to study analytically the evolution of the Universe for generic values of $s$.
Note that there is no simple closed formula of the modified Friedmann equation even in usual real LQC when one considers a regularization of the curvature operator with a spin $j$ greater than
$1/2$.  It is therefore not surprising that a similar phenomenon exists in the complex theory.

We also note that the Hamiltonian for the matter content is unchanged  compared to real LQC since it does not depend on the connection variable. One therefore recovers the standard expressions for the energy density as a function of the free scalar field as well as the Klein-Gordon equation for the dynamics of the free scalar field respectively given by
$$ \tilde{\rho}=\frac{1}{2}\dot\Phi{}^2  \;\;,\;\;\;\;\;\ddot\Phi+\frac{\dot{V}}{V}\dot\Phi=0.$$

\medskip

To have a better understanding of the dynamics in complex LQC, we proceed as in the standard real approach and we choose the smallest representation $s$ to regularize the Hamiltonian constraint.
We can refer to the same argument as in the real theory to explain this choice. Indeed, this choice corresponds to quantum cosmology states with the best possible coarse grained homogeneity. In the complex
theory, we have to choose $s=0$ in which case the edges of the corresponding spin-network states carry an area equal to the area gap $\Delta=2\pi \ell_p^2$ (\ref{complexagap}). 
Therefore, an edge color with $s=0$ is not equivalent to an edge not colored at all in the complex theory. 
With this choice the Hamiltonian constraint and the energy density simplifies according to the formulae:
\begin{eqnarray}
\tilde{H}= N(H_m - \tilde{\rho} V) \;\;\;\; \text{with} \;\;\;\; \tilde{\rho}=\frac{9}{8\pi G} \frac{\sinh^2(\lambda \tilde{b})}{\lambda^2}  f(\tilde{\theta}) \;\;\;, \;\;\;  f(\tilde{\theta})=\frac{ \tilde{\theta} \coth\tilde{\theta} -1}{\sinh^2 \tilde{\theta}}. \label{eq:rhotilde}
\end{eqnarray}
Contrary to the cases where $s\neq 0$, the energy density does not oscillate. Its dynamics is much simpler: it first increases until it reaches its maximal value $\tilde{\rho}_{max}$ then it decreases exponentially when 
$\tilde{b}$ (or $\tilde{\theta}$) becomes larger and larger. The shape of $\tilde{\rho}$ as a function of $\tilde{b}$ as derived from the r.h.s of Eq. (\ref{eq:rhotilde}) is depicted in the left panel of Fig. \ref{fig:complex} for $s=0$, clearly showing that is is bounded from above. To analyze further it is interesting to compute (and eventually to solve in some regimes) the dynamical equations satisfied by the physically relevant phase space parameters. It is easy to show that
the connection variable $\tilde{b}$ satisfies:
\begin{eqnarray}
\frac{\partial \tilde{b}}{\partial t} =  \{\tilde{b} , H_m - \tilde{\rho} V\} = - 4\pi G \left( \frac{\Pi^2}{2V^2} + \tilde{\rho}\right) \simeq -8\pi G \tilde{\rho}.
\end{eqnarray}
As $\tilde{\rho}>0$ when $s=0$, $\tilde{b}$ decreases when the cosmic time increases. In  other words, $\tilde{b}$ and $t$ go in opposite directions which means that the large $b$ and small $b$ asymptotic regimes correspond
respectively to the state of the Universe before and after the bounce. 

\begin{figure}
\begin{center}
	\includegraphics[scale=0.5]{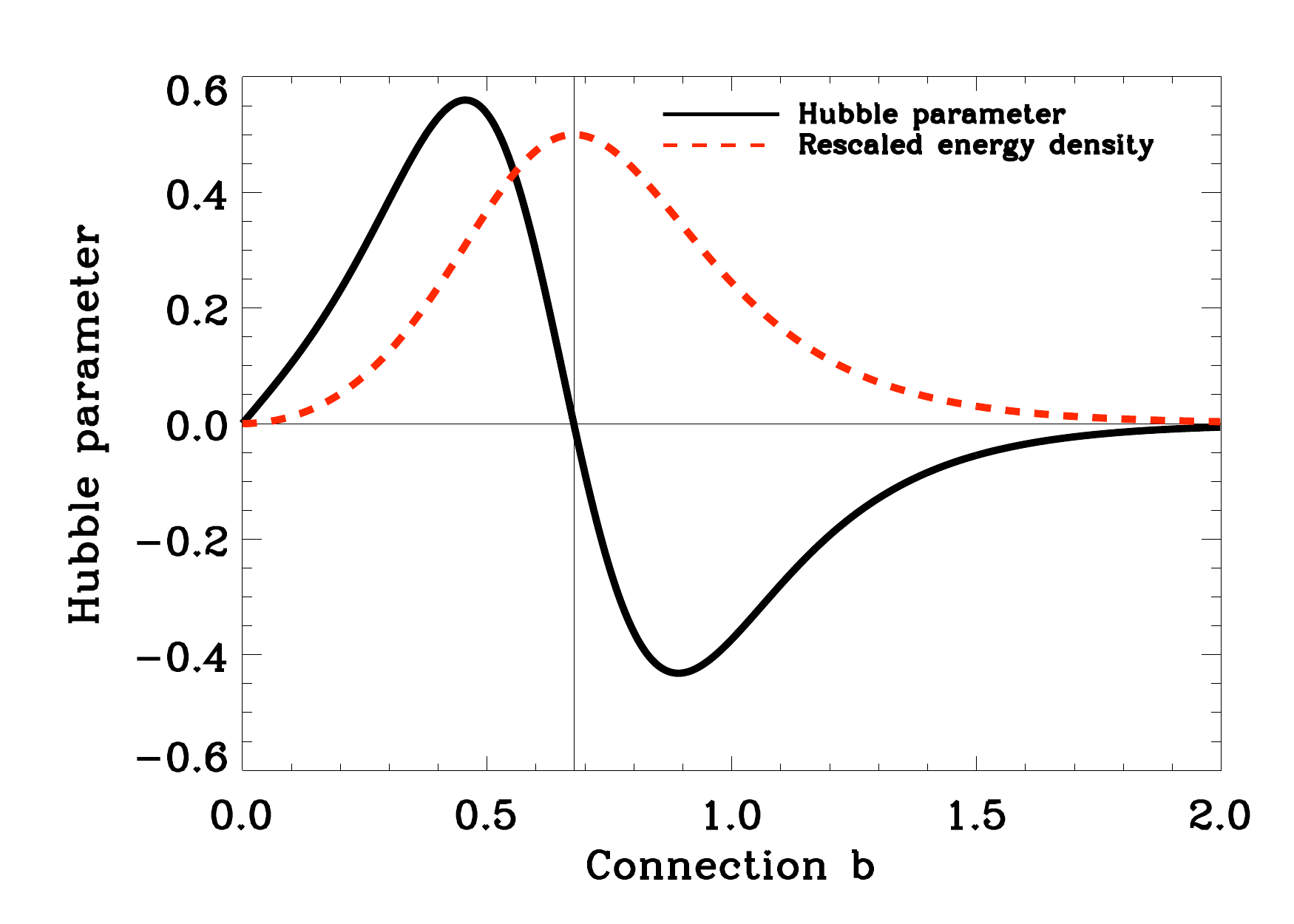}
	\caption{Hubble parameter (black curve) as a function of $\tilde{b}$ for $s=0$. The energy density (rescaled so as to be clearly displayed) is shown for comparison. The maximum of this energy density is indeed reached for a vanishing Hubble parameter.}
	\label{fig:hubble}
\end{center}
\end{figure}

It is also very interesting to compute the Hubble parameter (or equivalently the time evolution of the volume variable $V$ which is conjugate to $\tilde{b}$). Its expression  as a function of $\tilde{b}$ is given by:
\begin{equation}
	\frac{\dot{a}}{a}=\frac{\dot{V}}{3V}=\frac{1}{3}\left\{\tilde{\rho},V\right\}\Longrightarrow\frac{\dot{a}}{a}=\frac{4\pi G}{3}\frac{\partial\tilde\rho}{\partial\tilde{b}}.
\end{equation}
Using the equality (which follows immediately from (\ref{tildetheta}))
\begin{equation}
	\frac{\partial}{\partial\tilde{b}}= \lambda \frac{\sinh(\lambda \tilde{b})}{\cosh(\tilde{\theta}/2)} \, \frac{\partial}{\partial \tilde{\theta}}
\end{equation}
one arrives immediately at:
\begin{equation}
	\frac{\dot{a}}{a}=\frac{{3}}{2\lambda}\left[\sinh(2\lambda\tilde{b})\times f(\tilde\theta)+\frac{\sinh^3(\lambda\tilde{b})}{\cosh ({\tilde\theta}/{2})}\times\frac{\partial f}{\partial\tilde\theta}\right].
\end{equation}
This is displayed in Fig. \ref{fig:hubble}  as the black curve. The energy density is also shown as a dashed red curve, where it has been rescaled to be clearly displayed. This shows that the maximum of the energy density is indeed reach for a vanishing Hubble parameter, thus changing its sign at the bounce. Even if the singularity is removed and the Universe is bouncing when its volume achieves a minimal value, the precise dynamics is rather different than the real LQC dynamics.
First of all, the energy density is no more periodic as we have already said. 
Then the behaviors of the Universe before and after the bounce are very different and not symmetric as it is the case in usual LQC. This allows to make a clear distinction between
what happens before and after the bounce. For all these reasons, the dynamics of complex LQC seems to feature more interesting physical properties than the usual LQC. However, it is technically more involved to handle. 

To illustrate this different phenomenology, let us study the dynamics in the two asymptotic regimes $\tilde{b}\ll1$ and $\tilde{b}\gg1$. The corresponding asymptotic for the  energy densities have been computed previously (\ref{asympenergy}). 
For small values of $\tilde{b}$, the energy density is proportional to $\tilde{b}{}^2$, hence one recovers the standard Friedmann equation as expected. 

For large values of $\tilde{b}$, the phenomenology is more interesting. 
The energy density decreases exponentially with $\tilde{b}$ and this leads to a different version of the Friedmann equation. In that limit (keeping in mind that $\tilde{b}$ and $\tilde{\theta}$ are positive), 
the curvature variable $\tilde{\theta}$ and the ``connection" variable $\tilde{b}$ are  related by  
$$2 \exp\left( \frac{\tilde{\theta}}{2}\right) \sim \exp ( \lambda\tilde{b} ).$$ 
This relation enables us to write the energy density as a function of $\tilde{b}$ only from which
we can extract the Hubble parameter as follows:
\begin{equation}
	\tilde{\rho} \, \sim\, \frac{36}{\pi G \lambda}\, \tilde{b}~e^{-2\lambda\tilde{b}}\,\;\;\; \Longrightarrow \;\;\;\;
	\frac{\dot{a}}{a} \, \sim \, {-96} \, \tilde{b}~e^{-2\lambda\tilde{b}} \,.
\end{equation}
These two equations give the following modified Friedmann equation relating the``expansion" rate $(\dot{a}/a)$ of the Universe to the energy density $\tilde{\rho}$ of its content (here a massless scalar field) in the regime
where $\tilde{b} \gg 1$:
\begin{equation}
	\frac{\dot{a}}{a} \sim - \frac{8\pi G \lambda}{3}\times \tilde{\rho}. \label{eq:friedcomplex}
\end{equation}
In this regime, the Hubble parameter is therefore {\it linearly} related to the energy density. By squaring the above, one obtains 
$$\left(\frac{\dot{a}}{a}\right)^2= \frac{8\pi G}{3}\, \frac{{\tilde\rho}^2}{\rho_\lambda} \;\;\;\; \text{where} \;\;\;\; \rho_\lambda=\frac{3}{8\pi G \lambda^2}$$
is a constant with the dimension of an energy density ($\lambda^2$ has the same dimension as $G$), and which is of the order of few hundredths of the Planck energy density, $\rho_\lambda\simeq0.019\times\rho_\mathrm{Pl}$. This has to be compared to the standard Friedmann equation $(\dot{a}/a)^2=(8\pi G/3)\times {\tilde\rho}$. This shows that this branch $\tilde{b}\gg1$ corresponds to a low energetic universe ($\tilde\rho\to0$) but with a non-classical effective dynamics for its expansion rate. This differs from real LQC in which two asymptotically classical branches ($\lambda b\to0$ and $\lambda b\to\pi$) are connected via the bounce. Here, one connects a non-classical branch ($\tilde{b}\to+\infty$) for which the energy density is linearly related to the expansion rate, to an asymptotically classical branch ($\tilde{b}\to0$) for which one recovers the classical quadratic relation between the energy density and the expansion rate. We will discuss and propose physical interpretations for this results in the last section \ref{dis}.

For the moment, let us go further in the analysis of this regime by integrating the time evolution of the scalar field $\Phi(t)$ and the scale factor $a(t)$. This can be  solved exactly.
From the Klein-Gordon equation for the free massless field, one easily obtains 
$$\tilde\rho(t)=\tilde{\rho}_0\, \left(\frac{a_0}{a(t)}\right)^6$$
where $\tilde{\rho}_0$ and $a_0$ are the values of $\tilde{\rho}$ and $a$ at a given time $t_0$. 
Plugging this into the equation (\ref{eq:friedcomplex}), the modified Friedmann equation can be  solved as a function of the scale factor and we get:
\begin{equation}
	a(t)= a_0 \left[ 1 - 6 \frac{\tilde{\rho}_0}{\rho_\lambda} \frac{t-t_0}{\lambda} \right]^{1/6}
\end{equation}
where the time parameter is the cosmic time (from the beginning). Clearly, as time increases, the scale factor decreases and the energy density increases. We stress that the square root involved in the scale factor is always definite for $\tilde{b}\gg1$.The above-derived explicit solution for $a(t)$ is valid as long as the modified Friedmann equation (\ref{eq:friedcomplex}) is valid ({i.e.} $\tilde{b}\to\infty$), that is as long as $\tilde\rho\ll\rho_\lambda$. (We note that $\tilde\rho_0$ is assumed to be much smaller than $\rho_\lambda$, and that the relation $\rho\propto a^{-6}$ is valid on the entire range of values of $\tilde{b}$ since it only derives from the Klein-Gordon equation without any further assumption.) From the requirement that our approximation holds only for $\tilde\rho(t)\ll\rho_\lambda$, this fixes the time scale on which the above expression is valid to be:
$$
 t-t_0 \ll\frac{\lambda\rho_\lambda}{6\tilde\rho_0}\left[1- \frac{\tilde\rho_0}{\rho_\lambda} \right].
$$
Since $\tilde\rho_0/\rho_\lambda\ll1$, it is easily checked that the square root involved in $a(t)$ is positive-definite. Note that such an expression is valid only if $\tilde{b}\gg 1$, which means that the energy density must be very low and then scale factor large. 
When the scale factor in the previous expression becomes at the order of $1$ (or smaller), the previous expression is no more a good approximation. 

For completeness, we derive the expression of the free scalar field as a function of time. This is easily done using the expression of $\tilde{\rho}$ as a function of the scale factor:
\begin{equation}
	\Phi(t)=\Phi_0 + \frac{\lambda {\dot{\Phi}_0} }{3} \frac{\rho_\lambda}{\tilde{\rho}_0}\left\{1- \left[ 1 - 6 \frac{\tilde{\rho}_0}{\rho_\lambda} \frac{t-t_0}{\lambda} \right]^{1/2}\right\}
\end{equation}
where $\Phi_0$ and $\dot{\Phi}_0$ are the values of the field and its derivative at $t_0$.
For increasing time, the scalar field is as expected monotically increasing which justifies its use as a clock in this regime. 
As for the expression of the scale factor, such an expression 
is valid only when $\tilde{b}\gg 1$. 

\subsection{Exactly solvable complex LQC}
\begin{figure}
\begin{center}
	\includegraphics[scale=0.4]{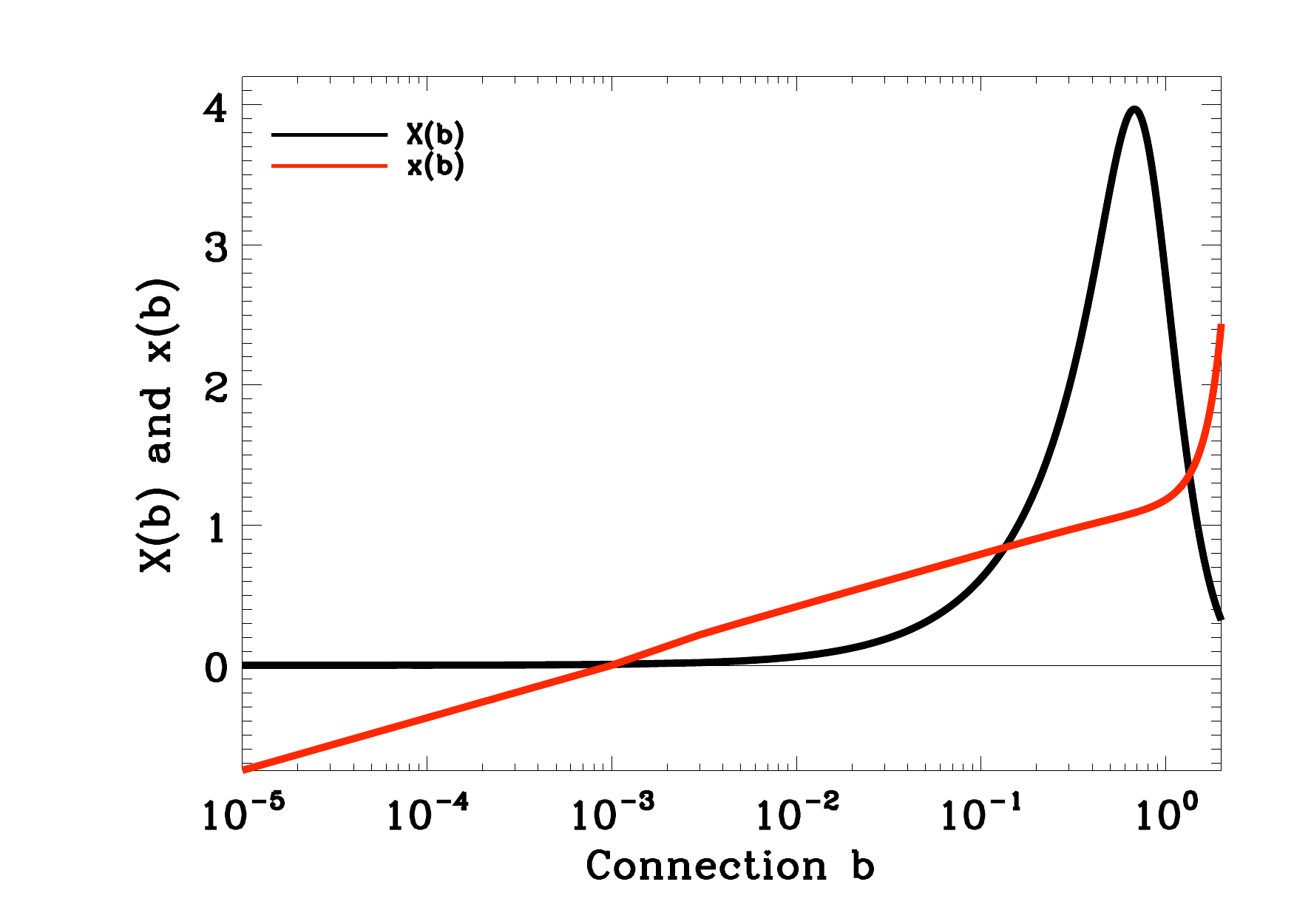}\includegraphics[scale=0.4]{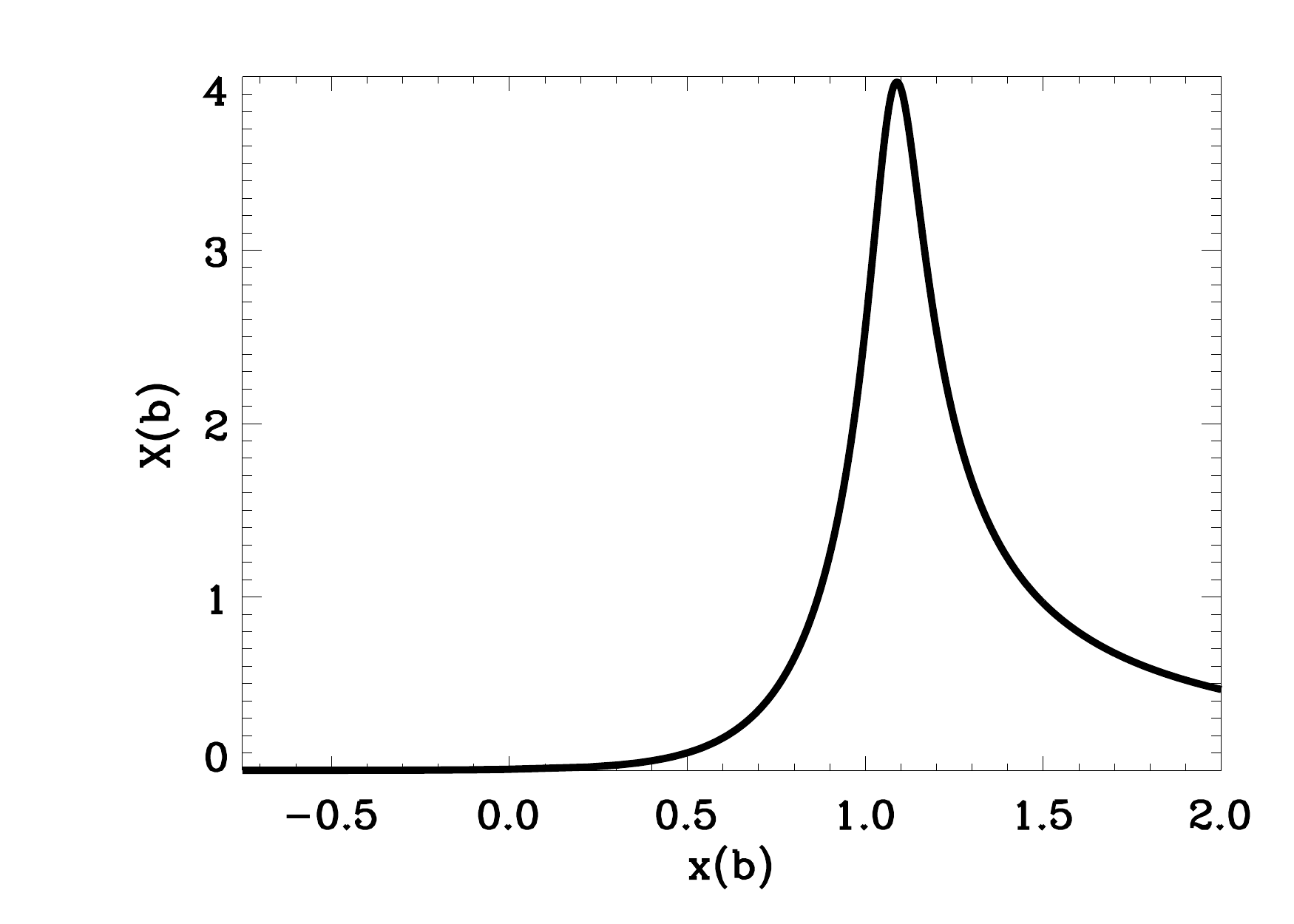}
	\caption{{\it Left panel:} $X$ as a function of $b$ (black curve) and $x$ as a function of $b$ (red curve). The connection $b_0$ is fixed to $10^{-3}$ for numerically computing the integral defining $x(b)$, meaning that $x(10^{-3})=0$. Because $b_0\ll1$, it is easily shown that for $b\ll1$, $x(b)\propto\log(b/b_0)$ for $b<b_0$. (Note that logarithmic scales are used for $b$.) {\it Right panel :} $X$ as a function of $x$, entering in the change of variables and in the expression of the expectation value for the volume operator. This function is always positive, admits a maximal value and falls off to zero when $x$ goes to $\pm \infty$}
	\label{xofb}
\end{center}
\end{figure}

To finish with the dynamics, we study the so-called exactly soluble  complex LQC. 
It is indeed well known that LQC can be solved exactly if one uses the scalar field as an internal clock  and 
also if one works in a suitable representation of the quantum algebra of operators. In this representation, the quantum states are functions of $\tilde{b}$ and the volume (or a related quantity) $V$ is an operator which acts
as a derivation. In other words, one chooses a polarization such that $\tilde{b}$ is the coordinate and $V$ the conjugate momentum. To make the exact resolution of the dynamics more concrete
we recall that the phase space is locally parametrized by the pairs of canonically conjugate variables
\begin{eqnarray}
\{\Pi,\Phi\}=1 \;\;\;\;\;\;\;\; \text{and} \;\;\;\;\;\;\; \{\tilde{b},v\}=1 \;\;\; \text{with} \;\;\; v=\frac{V}{4\pi G}.
\end{eqnarray}
Furthermore, in harmonic time (where $N=V=a^3$), when one expresses the Hamiltonian constraint with previous variables, one obtains immediately
\begin{eqnarray}
\tilde{H}= \frac{1}{2}\left( {\Pi^2} -\left( v X(\tilde{b})\right)^2 \right)= 0 \;\;\;\; \text{with} \;\;\;\;  X(\tilde{b})= 4\pi G \sqrt{2\tilde{\rho}}
\end{eqnarray}  
where $\tilde{\rho}$ is the energy density (for $s=0$) which is positive.
At the quantum level, the phase space variables become non-commutative operators and the Hamiltonian needs to an ordering prescription to be well defined. To define
the quantum constraint, we proceed  exactly as in real LQC. First we choose a polarization such that quantum states are wave functions $\chi(\tilde{b},\Phi)$ where $\Phi$ is viewed as an internal clock in the theory. Then, the 
ordering prescription is chosen in such a way that the action of the Hamiltonian  on a wave function $\Psi$ takes the form:
\begin{eqnarray}
\frac{\partial^2  }{\partial \Phi^2}\, \Psi({b},\Phi)  =  \left( X({b}) \frac{\partial}{\partial {{b}}} \right)^2  \Psi({b},\Phi)
\end{eqnarray}
where we omitted the tilda symbol \~{} on the variable $b$ to lighten the notations. 
This partial differential  equation can be easily recasted in the more compact form and can be written as a simple Klein-Gordon equation, thank to the following change of variables :
\begin{eqnarray}\label{KleinGordon}
\frac{\partial^2}{\partial \Phi^2}\,  \chi({x},\Phi) =  \frac{\partial^2}{\partial {{x}} ^2} \,  \chi({x},\Phi) \;\;\; \text{with}\;\;\;\; x(b) =  \int^{b}_{b_0} \; \frac{du}{X(u)}\,
\end{eqnarray}
where $b_0$ is a non relevant constant.
The wave functions $\chi$ and $\Psi$ are related by $\chi(x(b),\Phi)=\Psi(b,\Phi)$. Note that such a change of variables is possible because $X$ is a positive function and therefore $x(b)$ is a monotonic (increasing)
function of $b$.  Contrary to what happens  in real solvable LQC, the present change of variable  is not fully tractable in the sense that the integration defining $x(b)$ cannot be simplified and written in terms of elementary functions.
It is nonetheless easy to obtain equivalents of the function $x(b)$ in the regimes where $b \ll 1$ and $b \gg 1$:
\begin{eqnarray}
 \text{when} \, b \ll 1 & \text{then} & x(b) \sim  \frac{1}{\sqrt{{12 \pi G}}}\int_{b_0}^b \frac{du}{u} \sim \frac{1}{\sqrt{{12 \pi G}}} \log b \\
  \text{when} \, b \gg 1 & \text{then} & x(b) \sim \frac{1}{24} \sqrt{\frac{\lambda}{2\pi G}} \int_{b_0}^b du \, \frac{\exp( \lambda u)}{\sqrt{u}} \sim \frac{1}{24} \frac{1}{\sqrt{2\pi G \lambda}} \frac{\exp( \lambda b)}{\sqrt{b}}  .
\end{eqnarray}
Therefore $x(b)$ is an increasing function which runs from $-\infty$ to $+\infty$ as it should. The shape of the function $x(b)$ on the whole set $b \in ]-\infty,+\infty[$ is given in the left panel of Fig. (\ref{xofb}). From the asymptotic expressions of $x(b)$ and $\tilde\rho$ for $b \ll 1$ and $b \gg 1$, one easily derives the dependance of $X$ as a function of $x$ in these regimes to be (we remind that $X\propto\sqrt{\tilde\rho}$): $X\sim \exp(x)$ for $x\to-\infty$ ({\it i.e.} $b \ll 1$), and $X\sim 1/x$ for $x\to+\infty$ ({\it i.e.} $b \gg 1$).

Once we obtain the equation (\ref{KleinGordon}) for the wave function, we can proceed exactly as in usual real LQC \cite{LQCreview}. 
The invariance of the theory under inversion of the triad leads to the symmetry condition $\chi({x},\Phi) = - \chi({- x},\Phi)$ and then we can write the general solution of the Klein Gordon as
$  \chi({x},\Phi) =  (F(x_{+}) - F(x_{-}))/\sqrt{2}$.  The function $F$ is an arbitrary complex-valued function on the real line (constrained to satisfy some integration properties) and $x_{\pm} = \phi \pm x$ corresponds to the left/right moving coordinates.
The physical inner  product between two such states $\chi_1$ and $\chi_2$, associated to the functions $F_1$ and $F_2$, is the standard one:
\begin{eqnarray}
  \langle \chi_1 , \chi_2 \rangle= i \int_{-\infty}^{+\infty} dx \left( \partial_x \overline{F}_1 (x_+) F_2(x_+) -  \partial_x \overline{F}_1 (x_-) F_2(x_-) \right).
\end{eqnarray}

Following what is done in usual LQC, we can now evaluate the expectation value of the volume operator $\hat{V}=4\pi G \hat{v}$ on a state $\chi$ characterized by a function $F$.
In the $b$ representation, $\hat{v}= - i  \partial_b$ acts as a derivative operator and, due to the previous change of variables, its acts as follows:
\begin{eqnarray}
\hat{\nu} = - i  \frac{1}{X(x)} \partial_x
\end{eqnarray}
on functions of $x$.  As a consequence, we obtain immediately:
\begin{eqnarray}
 \langle \chi , \hat{V} \;  \chi  \rangle=  4\pi G  \int_{-\infty}^{+\infty} dx \, \lvert \frac{\partial F}{\partial x} \rvert^2 \left( \frac{1}{X(x-\phi)} + \frac{1}{X(\phi-x)}\right)
\end{eqnarray}
which converges when $\chi$ is a normalizable state. Finally, we obtain immediately that, whatever the normalizable state $\chi$ is, the expectation value of the volume operator is bounded from above according to
\begin{eqnarray}
  \langle \chi , \hat{V} \;  \chi  \rangle   \geq   \frac{ 8\pi G}{X_{max}}   \int_{-\infty}^{+\infty} dx \, \lvert \frac{\partial F}{\partial x} \rvert^2 
\end{eqnarray}
where $X_{max}$ is the maximum of $X$. This implies necessarily the existence of a bounce.   Furthermore, as in usual LQC, the expectation value of the volume tends to infinity when $\phi \rightarrow \pm \infty$.

\section{Discussion}
\label{dis}
Motivated by new ideas coming from black holes thermodynamics, we have constructed a new theory of Loop Quantum Cosmology where the Barbero-Immirzi parameter takes the complex value $\gamma=\pm i$. 
This construction relies on the analytic continuation of the regularized Hamiltonian constraint from real values of $\gamma$ to the complex value $\gamma=\pm i$. For the area operator to remain positive and
self-adjoint (i.e. with real and positive eigenvalues), it is furthermore necessary to change the representation $j$, which enters in the construction of the non-local curvature operator, into the complex number
$j=-1/2+is$ where $s \in \mathbb R$. This allowed us to define the Hamiltonian constraint for the new complex LQC that we studied mainly for $s=0$. Using first effective theory methods and then exactly solvable LQC 
techniques, we analyzed the dynamics in details. We showed that the bouncing scenario remains and the bounce relates two different branches of the Universe: a standard GR branch at the semi-classical limit (after the bounce)
and a new non ``classical" branch (before the bounce).  The latter phase corresponds to an Universe with law density but with high curvatures (because $b$ is large). Such Universes were also discovered in the totally different context
of brane cosmology \cite{Langlois} and it would be certainly very interesting to try to make a link between these two approaches if there exists any. On the other hand, it is obviously possible to transform the
unconventional Friedmann equation  (\ref{eq:friedcomplex}) into the usual one with a simple redefinition of the time variable $\tau$ (which satisfies $d\tau=\left({\tilde{\rho}}\right)^{1/2}dt$). Even if the time $\tau$
is no more the usual cosmic time in the standard GR branch, it could be possible that it plays such a role in the non-standard branch. We hope to study all these aspects in the future.

\medskip

To go further and understand more deeply this complex LQC, we need a  detailed study of the quantization and a precise construction of the kinematical and physical quantum states. 
In standard LQC, the expression of the non-local curvature operator (among other things) allows us to define the quantum states as trigonometric
functions of the variable $b$. In the complex theory, the suitable variable to work with seems to be the curvature $\tilde{\theta}$ rather than the connection variable $\tilde{b}$.  
Furthermore, the non-local curvature operator involves trigonometric and hyperbolic functions of $\tilde{\theta}$ and then it is expected that the quantum states would be
defined as sums of functions of the type $\exp( -m\tilde{\theta} + i n \tilde{\theta} )$. We hope to study all these aspects in details: construction of the scalar product, action
of the Hamiltonian constraint, robustness of the bouncing scenario, etc.  Such a model could help us understanding the quantization of the full complex theory of gravity and also 
finding our way towards the resolution of the reality conditions. It would be also very instructive to study the model in the framework of spin-foam models 
and to make a link with  Spin-Foam cosmology \cite{SFC}. This could open a new door towards the understanding of the dynamics in LQG.

\subsection*{Aknowledgments}
It is a pleasure to thank M. Geiller and  E. Wilson-Ewing for their comments and their help to improve the manuscript.

\end{document}